\documentclass[fleqn,10pt]{wlscirep}
\usepackage[utf8]{inputenc}
\usepackage[T1]{fontenc}
\usepackage{graphicx}

\newcommand{\e}{\varepsilon}

\newcommand{\s}{\sigma}
\newcommand{\G}{\Gamma}
\newcommand{\up}{\uparrow}
\newcommand{\down}{\downarrow}

\newcommand{\GFzub}[2]{\langle\!\langle #1|#2\rangle\!\rangle}

\newcommand{\exch}{\Delta \e_{\rm exch}}
\newcommand{\eM}{\Delta \e_{\rm M}}
\newcommand{\Pol}{\mathcal{P}}

\title{Spin-selective transport in a correlated double quantum dot-Majorana wire system}

\author[1,*]{Piotr Majek}
\author[1]{Ireneusz Weymann}
\affil[1]{Institute of Spintronics and Quantum Information, Faculty of Physics,
	Adam Mickiewicz University, ul. Uniwersytetu Pozna\'nskiego 2, 61-614 Pozna\'n, Poland}
\affil[*]{pmajek@amu.edu.pl}

\begin{abstract}
In this work we investigate the spin-dependent transport through
a double quantum dot embedded in a ferromagnetic tunnel junction
and side attached to a topological superconducting nanowire hosting Majorana zero-energy modes.
We focus on the transport regime
when the Majorana mode leaks into the double quantum dot competing
with the two-stage Kondo effect and the ferromagnetic-contact-induced exchange field.
In particular, we determine the system's spectral properties and analyze the
temperature dependence of the spin-resolved linear conductance
by means of the numerical renormalization group method.
Our study reveals unique signatures of the interplay between the spin-resolved tunneling, the Kondo effect
and the Majorana modes, which are visible in the transport characteristics.
In particular, we uncover a competing character
of the coupling to topological superconductor and that to ferromagnetic leads,
which can be observed already for very low spin polarization of the electrodes.
This is signaled by an almost complete quenching of the conductance in one of the spin channels
which is revealed through perfect conductance spin polarization.
Moreover, we show that the conductance spin polarization can change sign depending on the magnitude
of spin imbalance in the leads and strength of interaction with topological wire.
Thus, our work demonstrates that even minuscule
spin polarization of tunneling processes can have
large impact on the transport properties of the system.
\end{abstract}

\begin{document}


\flushbottom
\maketitle

\thispagestyle{empty}

\section*{Introduction}

The exploitation of spin degrees of freedom in studies of transport through nanostructures
has opened new avenues for exploring a plethora of different many-body phenomena 
where the interplay of spin and charge determines the system's transport behavior
\cite{Wolf2001,DasSarma2004,Barnas2008Sep,Wiesendanger2009Nov,Awschalom2013Mar,Hirohata2020Sep}.
Interestingly, the spin-selective investigations have allowed one to explore
various bound states, including topologically protected states, such as the Majorana 
zero-energy modes \cite{Sticlet2012Mar,He2014Jan,Kotetes2015Nov,Chirla2016Jul,Jeon2017Nov,
	Hoffman2017Jul,Li2018Mar,Rancic2019Apr,Schuray2020Jul,Wang2021Feb,Bjerlin2022Jul,Huguet2023Oct}.
The Majorana bound states are solid-state realizations of Majorana fermions,
i.e., particles that are their own anti-particles \cite{Majorana1937Apr}.
The Majorana modes have been predicted to appear
in a topologically nontrivial phase of a spinless tight-binding chain with superconducting
pairing of the same spin species between neighboring sites, i.e., in the so-called Kitaev chain \cite{Kitaev2001}.
Experimentally, such a chain, also referred to as a Majorana wire, can be implemented by
placing a semiconducting nanowire with strong spin-orbit interaction
in proximity with an $s$-wave superconductor
in the presence of external magnetic field \cite{Lutchyn2010Aug,Sau2010Jan},
or in a chain of magnetic adatoms on superconducting substrate \cite{Schneider2022Apr,Wiesendanger_2023,Schneider2023May}.
Moreover, Kitaev chains could be also
implemented in the absence of external magnetic fields, by 
exploiting spiral spin structure of the chain atoms~\cite{Maska2021Jun}.
Very recently, a minimal Kitaev chain based on coupled quantum dots has also been implemented
\cite{Dvir2023Feb,Bordin2023Sep,Bordin2024Feb}.

Although a great experimental endeavor has already been taken
\cite{Mourik2012May,Deng2012Nov,Das2012Nov,Albrecht2016Mar,Gul2018Jan},
which has been stimulated by possible applications in fault-tolerant quantum computation
\cite{Kitaev2003Jan,Nayak2008Sep,BraidingReview},
a unique and unambiguous confirmation of the existence of Majorana modes 
is still awaited, despite great recent progress \cite{Dvir2023Feb,Quantum2023Jun}.
From this perspective, it is important to continue
further investigations on transport properties that could provide 
other indications of the presence of Majorana quasiparticles \cite{Alicea2012Jun,Aguado2017Oct,Lutchyn2018May,Zhang2019Nov,Flensberg2021Oct}.
One of promising ways is to explore the transport characteristics
of side-attached zero-dimensional systems, to which topologically protected
Majorana modes can leak giving rise to fractional values of the conductance
\cite{Vernek2014Apr,Ruiz-Tijerina2015Mar,Deng2016Dec,Deng2018Aug}.
In this respect, there have already been examinations
involving both single and multiple quantum dot systems \cite{Liu2011Nov,Leijnse2011Oct,Cao2012Sep,Gong2014Jun,
	Liu2015Feb,Liu2017Aug,Prada2017Aug,Ptok2017Nov,Gorski2018Oct,Cifuentes2019Aug,
	Ricco2019Apr,Wang2019May,Zienkiewicz2019Oct,Chen2020Jul,
	Wrzesniewski2021Mar,Feng2022Jan,Majek2022May,Liu2023Aug,Diniz2023Jan}.
Moreover, the considerations have also concerned the strong coupling regime,
where Majorana quasiparticles interact with the Kondo correlations \cite{Golub2011Oct,Lee2013Jun,Cheng2014Sep,Silva2020Feb}.
As mentioned above, further information about Majorana modes
could be obtained from spin-selective investigations, e.g. by embedding
the nanostructure into a ferromagnetic junction or by using spin-polarized spectroscopy.
Such considerations have been recently performed for single quantum dots
attached to topological superconducting wires
\cite{Weymann2017Apr,Weymann2017Jan},
while much less is known about the properties of double quantum dot systems,
in which the Majorana-Kondo interplay results in further interesting features \cite{Weymann2020Jun,Majek2021Aug}.
The goal of this work is therefore to extend the existing studies
by addressing the problem of spin-dependent 
transport properties of a double quantum dot system attached 
to nanowire hosing Majorana quasiparticles at its ends.
In particular, we consider the case when one of the dots is coupled
to ferromagnetic electrodes, while the second quantum dot is connected
to the Majorana wire forming a T-shaped geometry, as displayed in Fig.~\ref{fig:model}.
Our considerations are performed with the aid of the numerical renormalization
group (NRG) method \cite{Wilson1975,NRG_code}, which allows for taking into account
all correlation effects in a fully non-perturbative manner \cite{Bulla2008}.
We have in particular determined the spectral properties of the system
for various coupling strengths to the Majorana wire.
Moreover, we have analyzed the temperature as well as gate voltage
dependence of the spin-resolved linear conductance.
These quantities showed features of the interplay between 
an exchange field, that is triggered by the presence of ferromagnets \cite{Martinek2003,Martinek2005,Gaass2011},
the Kondo correlations\cite{Kondo1964,Goldhaber1998} and the Majorana quasiparticles.
In particular, we demonstrate that, depending on the strength of coupling to the Majorana wire,
the system's conductance strongly depends on the value of spin polarization
of ferromagnetic contacts, and already very low spin polarization
is sufficient to drastically modify the conductance of the device.
Our work provides thus further insight into the 
physics of hybrid Majorana wire-quantum dot systems
where spin-selective transport plays an important role.

\section*{Methods}

\subsection*{System's Hamiltonian}

\begin{figure}[t]
	\centering
	\includegraphics[width=0.5\linewidth]{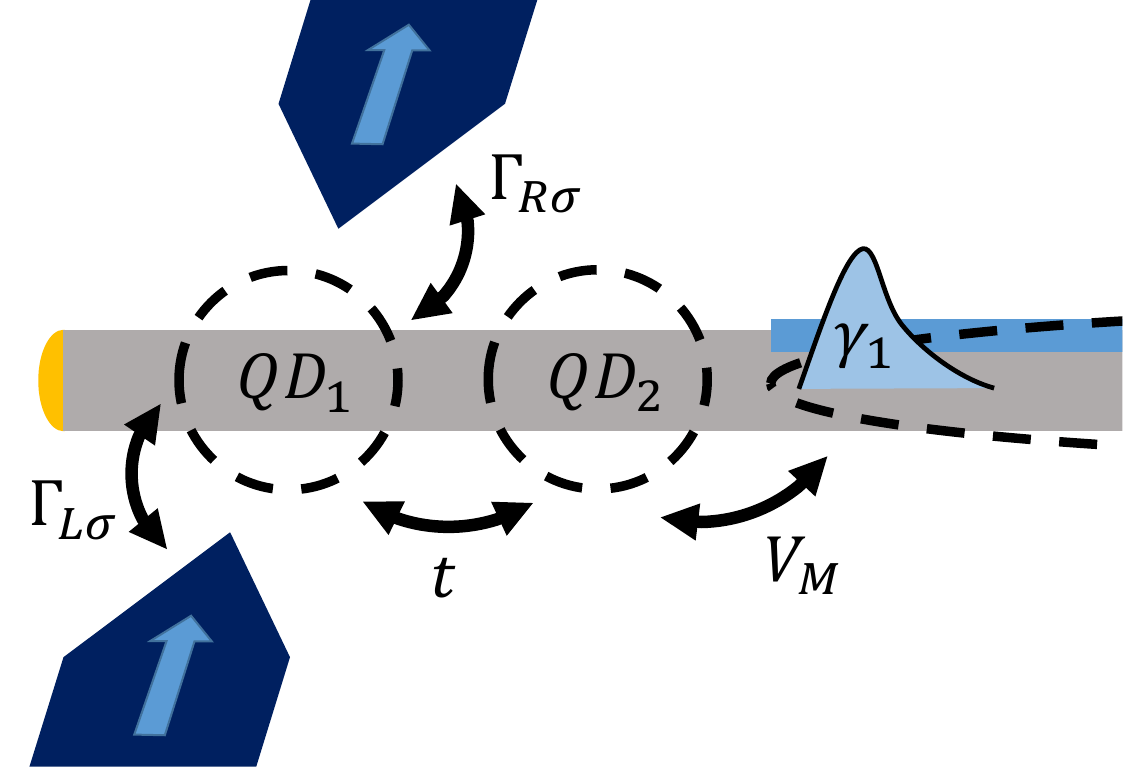}
	\caption{The schematic of the considered double quantum dot-Majorana wire system.
		The first quantum dot is attached to two ferromagnetic leads through the coupling strengths $\Gamma_{j\s}$,
		the two dots are coupled through the hopping matrix elements $t$,
		while the second quantum dot is connected to a superconducting nanowire
		in the topological phase through the hopping matrix elements $V_M$.
		$\gamma_1$ denotes the Majorana zero-energy mode that emerges
		at the end of the nanowire.}
	\label{fig:model}
\end{figure}

The considered system is presented in Fig.~\ref{fig:model}. It consists of  
two quantum dots (QD), arranged in a T-shaped geometry, attached to a 
topological superconducting nanowire hosting Majorana bound states at its 
ends. The first quantum dot (QD1) is coupled to the external ferromagnetic 
leads and the second one (QD2) to the nanowire. The general Hamiltonian that 
models the system can be written as $H = H_{\rm leads} + H_{\rm tun} + H_{\rm 
	DD-Maj}$. The first part,
\begin{equation}
	H_{\rm leads} = \sum_{r=L,R}\sum_{\mathbf{k}\sigma}
	\varepsilon_{r\mathbf{k}\sigma} c^\dag_{r\mathbf{k}\sigma} c_{r\mathbf{k}\sigma},
\end{equation}
describes the leads as a source of non-interacting electrons with
energy $\varepsilon_{r\mathbf{k}\sigma}$ in the left and right ($r = L, R$, 
respectively) electrode. The operator $c^\dag_{r\mathbf{k}\sigma}$ creates an 
electron with spin $\sigma = \up, \down$ and momentum $\mathbf{k}$ in the 
$r$-th electrode.
The next part of the Hamiltonian models the tunneling between the first quantum 
dot
and the electrodes
\begin{equation}
	H_{\rm tun} = \sum_{r=L,R}\sum_{\mathbf{k}\sigma} v_{r \s} 
	\left(d^\dag_{1\sigma}
	c_{r\mathbf{k}\sigma} + c^\dag_{r\mathbf{k}\sigma} d_{1\sigma} \right),
\end{equation}
where $v_{r \s}$ describes the momentum independent tunnel matrix elements, and 
$d^\dag_{1\sigma}$ creates an electron with spin $\sigma$ on the first quantum 
dot, which is directly attached to the ferromagnetic leads. The last term of 
the Hamiltonian models effectively the double quantum dot system
coupled to the superconducting topological nanowire.
It can be written as \cite{Flensberg2010Nov,Liu2011Nov}
\begin{equation}
	H_{\rm DD-Maj}=\sum_{j=1,2}\sum_{\sigma} \varepsilon_j d_{j\sigma}^\dag 
	d_{j\sigma}
	+ U \sum_{j=1,2} d_{j\uparrow}^\dag d_{j\uparrow} d_{j\downarrow}^\dag 
	d_{j\downarrow}
	+ t \sum_\sigma (d_{1\sigma}^\dag d_{2\sigma} +  d_{2\sigma}^\dag 
	d_{1\sigma})
	+ \sqrt{2} V_M (d^\dag_{2\downarrow} \gamma_1 + \gamma_1 d_{2\downarrow}) ,
\end{equation}
where the first two terms describe the first and second ($j = 1, 2$) quantum dot
with energy $\varepsilon_j$ and Coulomb correlations $U$,
assumed to be equal for both dots, respectively.
The matrix element $t$ stands for the hopping between the dots.
The last part couples the spin-down electrons on the second quantum dot
with the Majorana bound state $\gamma_1$ at the end of the topological 
superconducting nanowire,
with $V_M$ being the relevant tunnel matrix elements.
Here, we assume that the wire is sufficiently long, such that
the overlap between the Majorana modes at both ends of the nanowire is negligible.
The broadening of the first quantum dot energy level due to the coupling to
ferromagnetic leads is given by $\G_{r \s} = \pi \rho_{r \s} v_{r \s}^2$, 
where $\rho_{r \s}$ stands for the density of states at the Fermi level of the 
lead $r$ for spin $\s$. This coupling can be rewritten as $\G_{r \s} = (1 + \s 
p_r) \G_r$, with $\G_r = (\G_{r \up} + \G_{r \down})/2$ and $p_r$ referring to 
the spin polarization of the lead $r$. 
We assume a flat density of states of the leads and use its half-width as the energy unit,  $D \equiv 1$. 
Moreover, for the sake of simplicity of further analysis,
we assume that the quantization axis coincides with 
the Majorana polarization and that Majorana mode couples
to spin-down electrons on the second dot \cite{Lee2013Jun,Weymann2017Apr,Weymann2020Jun}.
In this work we are interested in the linear response transport properties.
Then, one can perform an orthogonal left-right transformation
to a new basis, in which the first quantum dot couples
only to an even linear combination of the left and right electron fields,
with a new coupling strength $\Gamma = \G_L + \G_R$ and an average spin polarization $p=(p_L + p_R)/2$.
Because such a transformation can be also performed for systems
with asymmetric couplings \cite{Wojcik2013Jan}, our considerations shall be relevant 
for another geometries as well, e.g. the ones involving only one ferromagnetic contact,
which could be a substrate or a spin-polarized tip of STM.

\subsection*{Computation method}

The calculations are performed with the aid of the numerical renormalization
group method involving the full density matrix \cite{Bulla2008,NRG_code,Andreas_broadening2007}.
The clue of the NRG method is a logarithmic discretization
of the conduction band and mapping of such discrete Hamitlonian
to a tight-binding chain with exponentially decaying hopping integrals $\xi_n$ \cite{Wilson1975}.
In our case the NRG Hamiltonian has the following form
\begin{equation}
	H = H_{\rm DD-Maj} + \sum_{\sigma} \sqrt{\frac{2\Gamma_\sigma}{\pi}}
	\left(d^\dag_{1\sigma} f_{0\sigma} + f^\dag_{0\sigma} d_{1\sigma} \right)
	+\sum_{n,\sigma} \xi_n  
	\left(f^\dag_{n\sigma} f_{n+1\sigma} + f^\dag_{n+1\sigma} f_{n\sigma} \right),
\end{equation}
where $f_{n\sigma}$ is the annihilation operator of an electron with spin $\sigma$ on the $n$th site of the chain.
This Hamiltonian can be solved iteratively by keeping an appropriate number
of low-energy eigenstates during the iteration $N_{\mathrm{kept}}$. In our computations we keep at least
$N_{\mathrm{kept}} = 4000$ states. We also use the 
logarithmic discretization parameter $\Lambda = 2$.
The spectral data is determined in the Lehmann representation
and collected in logarithmic bins that are then broadened
to obtain smooth functions. On the other hand,
the conductance is obtained directly from the discrete
data without the need to introduce broadening \cite{Weymann2013Aug}. 
In calculations, we make use of the conservation of spin-up particles
and also exploit the parity symmetry of the Hamiltonian.

\section*{Results and discussion}
\subsection*{Relevant energy scales}

The transport properties of the system are conditioned by relative interplay
of various energy scales. One of such scales, referred to as the exchange field, is set by the 
spin-resolved charge fluctuations between the first quantum dot and ferromagnetic contacts.
Such exchange field can be estimated from the second-order perturbation theory. 
Assuming zero temperature and vanishing hopping between the dots, one then finds \cite{Martinek2003}
\begin{equation}
	\exch \approx \frac{2p \Gamma}{\pi} \log \left| \frac{\e_1}{\e_1+U}\right|.
\end{equation}
As can be seen from the above formula, $\exch$ vanishes for 
$\e_1 = -U/2$. Thus, finite splitting of the dot level occurs
only when there is a detuning from the particle-hole symmetry point.
Another spin splitting can be also induced by the coupling to topological superconductor,
which in the leading order in $V_M$ can be written as~\cite{Lee2013Jun,Weymann2017Apr}
\begin{equation}
	\eM \approx \frac{2V_M^2}{U^2} \left(2\e_2+U\right),
\end{equation}
again, this splitting is only finite for $\e_2\neq -U/2$. Nevertheless,
it is important to note that even for $\e_2= -U/2$, the presence of Majorana quasiparticles
can give rise to quantum interference effects
that could result in fractional values of the conductance \cite{Lee2013Jun}.
The relevant energy scale associated with such Majorana interference
will be denoted by $\Gamma_M$.
We also note that the above formulas are only approximate
and one needs to keep in mind that, due to the hopping
between the two dots, both splittings will vanish
only when the system is fully at the particle-hole symmetry point,
i.e. when $\e_1 = \e_2 = -U/2$.

Further important energy scales are established by the strong electron correlations driving the Kondo effect.
In T-shaped double quantum dots a two-stage Kondo effect 
can emerge \cite{Pustilnik2001Nov,Vojta2002Apr,Cornaglia2005Feb,Chung2008Jan,Sasaki2009Dec,Wojcik2015}.
In this type of the Kondo phenomenon, with lowering the temperature $T$,
the spin of the first quantum dot becomes screened when $T<T_K$,
giving rise to an enhanced conductance.
However, further decrease in $T$,
gives rise to screening of the spin of the second quantum dot,
which happens for $T<T^*$, where $T^*$ denotes the second-stage
Kondo temperature. Then, since the second dot is not directly
coupled to the leads, the conductance becomes suppressed.
In the case of ferromagnetic leads,
the Kondo effect will develop in the system only when the Kondo temperature
$T_K$ is larger than the exchange field, $T_K> |\exch|$, which 
will happen around $\e_1\approx -U/2$.
Thus, $T_K$ for $\e_1 =  -U/2$ could be estimated from \cite{Haldane1978Feb,Martinek2003}
\begin{equation}
	T_K\approx \sqrt{\frac{U\Gamma}{2}}\exp\left[ -\frac{\pi U \;{\rm arctanh} (p)}{8\Gamma\; p}  \right].
\end{equation}
On the other hand, the second-stage Kondo temperature at the particle-hole symmetry point
can be found from \cite{Pustilnik2001Nov,Cornaglia2005Feb}
\begin{equation}
	T^* \approx a \; T_K \; \exp\left( \! -b \; \frac{ T_K U}{4t^2}\right),
\end{equation}
where $a$ and $b$ are constants of the order of unity \cite{Wojcik2015Apr}.

\subsection*{Spectral properties}

\begin{figure}[t]
	\centering
	\includegraphics[width=1\linewidth]{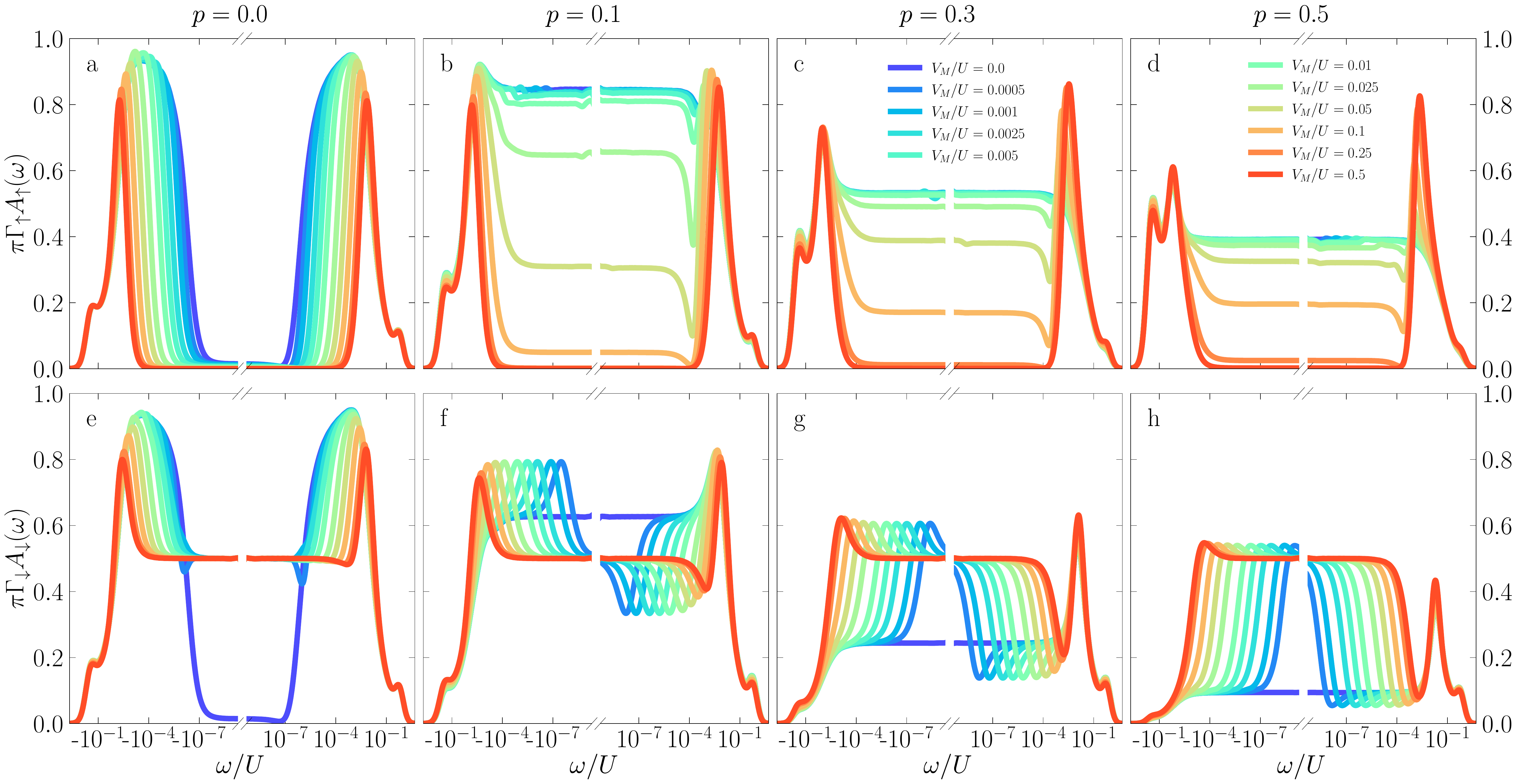}
	\caption{The normalized spin-resolved spectral functions
		of the first quantum dot plotted as a function of energy $\omega$ on the logarithmic scale.
		The top row (a-d) presents the spin-up contribution,
		while the bottom row (e-h) shows the spin-down contribution to the
		spectral function calculated for different values of the coupling
		to the topological superconductor  $V_M$, as indicated in the legend. 
		The consecutive columns present the results for different values of the 
		spin polarization of the leads $p$, as indicated. 
		The other parameters are as follows: $U=0.2$,
		$\Gamma=0.02$, $t=0.004$, in units of band half-width, and $T=0$.
		The quantum dots' orbital levels are set as
		$\varepsilon_{1} = -U/3$ and $\varepsilon_{2} = -U/2$.}
	\label{fig:fig2}
\end{figure}

Before analyzing the system's conductance, it is important to discuss the behavior of the 
spin-resolved spectral functions that will determine the transport characteristics.
The spin-resolved spectral function of the $j$th quantum dot is defined as,
${A_{j\sigma} (\omega) = - (1/\pi) \;\mathrm{Im} \GFzub{d_{j \s}}{d^{\dagger}_{j 
\sigma}}_\omega^\mathrm{ret}}$, where $\GFzub{d_{j \s}}{d^{\dagger}_{j 
\sigma}}_\omega^\mathrm{ret}$ is the Fourier transform of the retarded Green's 
function of the corresponding quantum dot.
In the following, we will focus on the behavior of $A_{1\sigma} (\omega) \equiv A_{\sigma} (\omega)$,
since it determines the system's conductance in the considered geometry, see Fig.~\ref{fig:model}.
We also notice that the case of fully particle-hole symmetric model
would essentially correspond to a nonmagnetic system with modified Kondo temperature \cite{Weymann2020Jun}.
Therefore, we will consider the cases when at least one of the quantum dots' energy levels is detuned from 
its particle-hole symmetry point, such that there is always finite splitting associated either with Majorana mode
or with proximity to ferromagnetic leads.

Figure~\ref{fig:fig2} presents the energy dependence of the spin-resolved normalized spectral function
of the first quantum dot, $\pi \Gamma_\sigma A_\sigma(\omega)$.
To resolve the relevant energy scales, the spectral functions
are plotted on a logarithmic scale, and we consider the case of zero temperature.
The top row displays the spin-up component,
while the bottom row shows the spin-down contribution to the spectral function.
Each curve is calculated for different coupling $V_M$ to the superconducting
nanowire, as indicated in the legend,
while each column refers to a different value of the spin polarization of the leads $p$,
as marked in the figure.
This figure is calculated for $\varepsilon_{1} = -U/3$ and $\varepsilon_{2} = -U/2$,
such that $\exch\neq 0 $, while $\eM \approx 0$.

The first column presents the reference case when $p = 0$,
i.e. when the electrodes are nonmagnetic.
Here, one can distinguish three different energy scales:
the ones associated with the first and the second stage of the Kondo screening,
$T_K$ and $T^*$, respectively, 
and the third one related to the quantum interference with the Majorana wire,
denoted by $\Gamma_M$.
As can be seen in the figure,
for $V_M=0$, the spectral function exhibits two resonances 
around $\omega \approx \pm T_K$ due to the first-stage Kondo effect,
while for $|\omega| \lesssim T^*$, $A_{\sigma}(\omega)$
becomes suppressed due to the second-stage of Kondo screening.
Finite coupling to Majorana wire, generally
blocks the second-stage screening in the spin-down channel,
giving rise to $\pi\Gamma_\down A_\down(0) = 1/2$.
The range of energies around the Fermi level
where $\pi\Gamma_\up A_\up (\omega) = 0$ while $\pi\Gamma_\down A_\down(\omega) = 1/2$
grows with increasing $V_M$, which indicates that 
the presence of Majorana mode alters the 
Kondo correlations down to energy scales of $\omega \approx \Gamma_M$.
In other words, finite $V_M$ enhances the second stage Kondo temperature.
We note that qualitatively similar results are obtained
even for finite $p$ when $\varepsilon_{1} = \varepsilon_{2} = -U/2$.

\begin{figure}[t]
	\centering
	\includegraphics[width=1\linewidth]{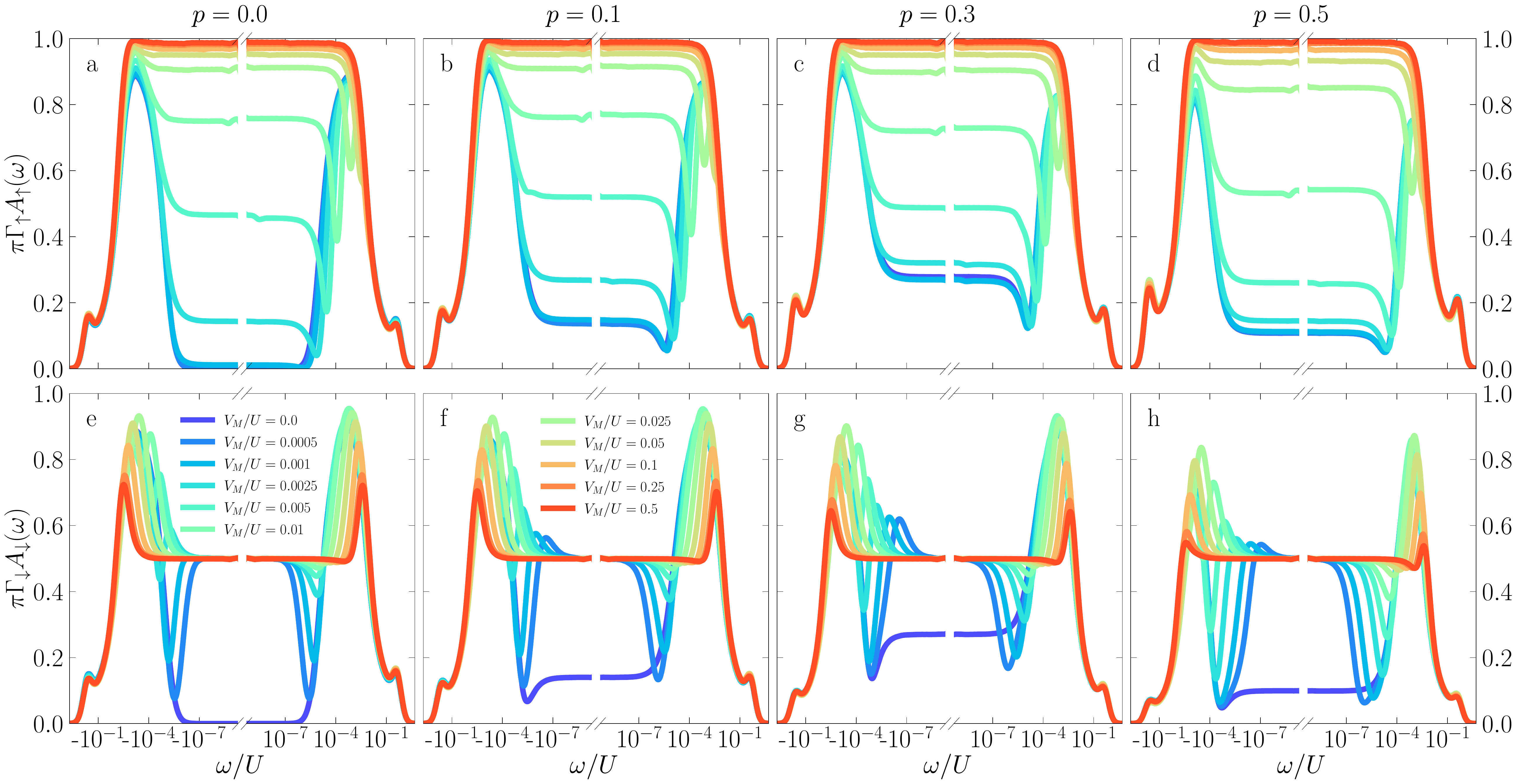}
	\caption{The normalized spin-resolved spectral functions
		calculated for $\varepsilon_{1} = -U/2$ and $\varepsilon_{2} = -U/3$.
		The other parameters are the same as in Fig.~\ref{fig:fig2}.}
	\label{fig:fig3}
\end{figure}

\begin{figure}[t]
	\centering
	\includegraphics[width=1\linewidth]{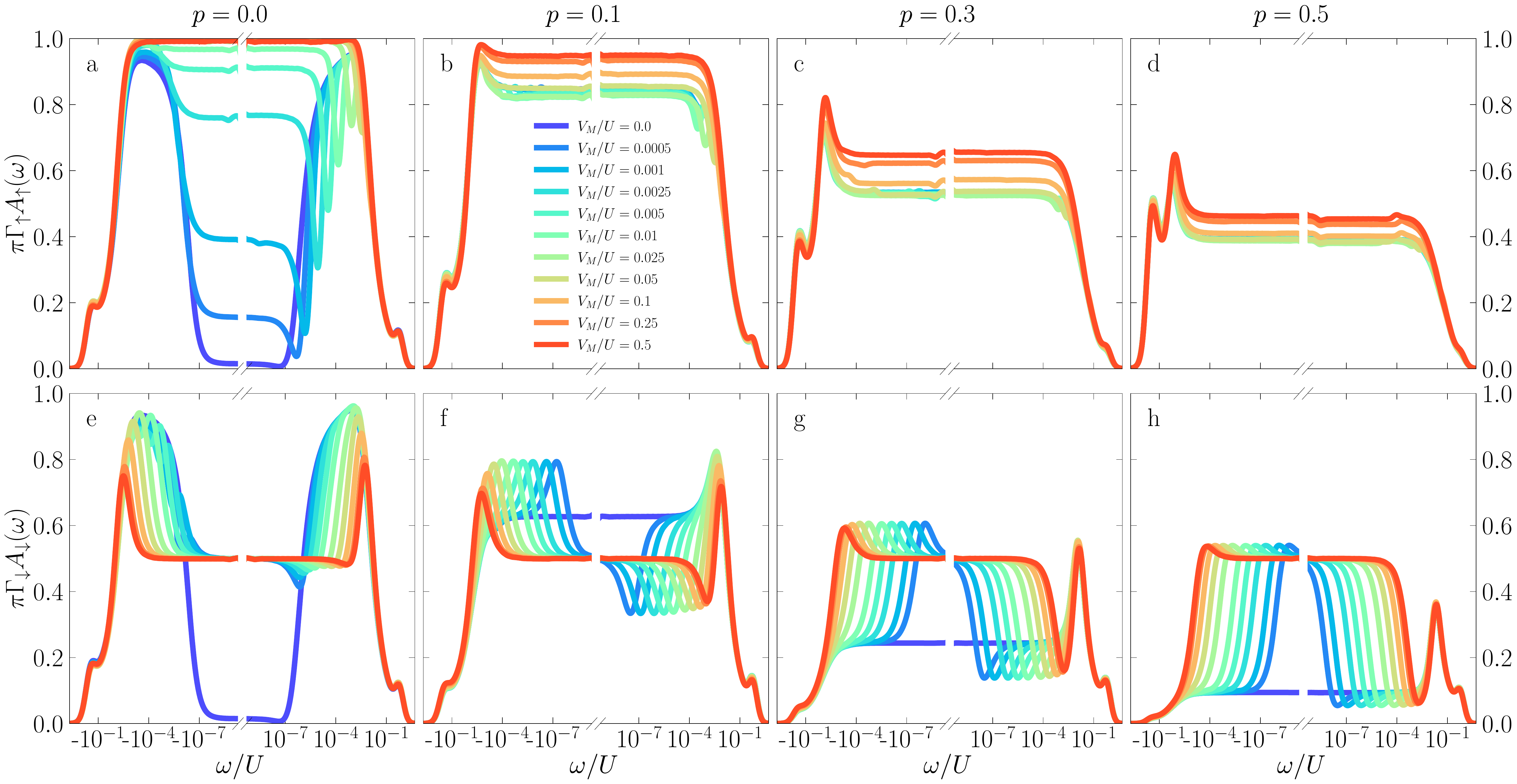}
	\caption{The normalized spin-resolved spectral functions
		calculated for $\varepsilon_{1} = \varepsilon_{2} = -U/3$
		and the other parameters are the same as in Fig.~\ref{fig:fig2}.}
	\label{fig:fig4}
\end{figure}

The cases of finite spin polarization are shown in the next columns
of Fig.~\ref{fig:fig2}, where the second column is calculated for $p=0.1$,
the third one for $p=0.3$, while the last one displays the case of $p=0.5$.
Note that this sequence also applies to Figs.~\ref{fig:fig3} and \ref{fig:fig4}.
As can be seen, having non-zero leads polarization $p$,
which results in finite exchange field $\exch$,
significantly modifies the whole picture discussed above for both spin contributions.
When the spin polarization is introduced, 
the majority contribution is associated with the spin-up electrons.
As can be seen in Fig.~\ref{fig:fig2}(b), this suppresses the second stage of Kondo 
screening for $V_M = 0$, restoring the low energy transport.
This happens since $T^*\lesssim |\exch| \lesssim T_K$.
Increasing the coupling to the Majorana wire, the influence of the ferromagnetic leads is 
being reduced (since $T^*$ grows with $V_M$),
and the system demonstrates the two-stage Kondo effect again once $V_M\gtrsim 0.1\;U$,
resulting in $A_\uparrow(0) = 0$.
One can also notice a dip that occurs for $\omega \approx 10^{-4}\;U$,
which deepens as the coupling to topological superconductor reaches $V_M=0.01\;U$.
This deep is an indication of the presence of exchange field splitting of the first quantum dot level.
On the other hand, the spin-down normalized spectral function
$\pi \Gamma_\downarrow A_\downarrow (\omega)$ for $p=0.1$ is shown in Fig. \ref{fig:fig2}(f).
As for the spin-up contribution, when there is no coupling to the Majorana wire, 
the presence of ferromagnetic leads affects
the spin screening on the second quantum dot, what 
results in destroying the second-stage of the Kondo effect.
However, when the Majorana coupling is present in the system, it immediately restores the 
low-energy spectral function to the value of $\pi \Gamma_\downarrow A_\down (0) = 1/2$.
Interestingly, one can also observe asymmetric peaks and dips,
which maximum (minimum) shifts with the increasing value of coupling
to Majorana wire. These features occur at energies corresponding to $\Gamma_M$.

This picture is modified for the higher values of the leads 
polarization, as shown in further columns of Fig.~\ref{fig:fig2}.
Generally, one observes a gradual suppression of the 
low-energy spectral functions in both spin components.
This is due to the fact that for the considered values of $p$, 
$T_K\lesssim |\exch|$ for $V_M=0$, such that also the first stage Kondo effect becomes suppressed. 
Nonetheless, for finite coupling to Majorana wire,
one can observe a restoration of the second-stage screening,
which results in $A_\uparrow (0) = 0$, while 
for the spin-down component one again finds,
$\pi \Gamma_\downarrow A_\downarrow (0) = 1/2$.
What also draws the attention here is
the suppression of asymmetric minima and maxima
visible in $A_\downarrow (\omega)$ with increasing $p$.
When the exchange field is considerable, 
see Fig.~\ref{fig:fig2}(h),
one basically observes a restored resonance 
at the Fermi level with 
$\pi \Gamma_\downarrow A_\downarrow (0) = 1/2$.

The case when the quantum dots' orbital levels
are tuned to $\varepsilon_{1} = -U/2$ and $\varepsilon_{2} = -U/3$ is shown in Fig.~\ref{fig:fig3}.
(Note that this figure is calculated for the same set of parameters as Fig.~\ref{fig:fig2}.)
This corresponds to the situation when there is ferromagnet-induced exchange field is negligible
$\exch\approx 0$, while there is a finite splitting caused by the coupling to the Majorana wire, $\eM\neq 0$.
In the first column the case of nonmagnetic leads is presented.
Contrary to the situation shown in Fig.~\ref{fig:fig2},
increasing the coupling to the topological superconductor,
the Kondo peak starts to form, reaching its maximum when $V_M \gtrsim 0.1\;U$.
The spin-down spectral function exhibits a weaker influence 
on the coupling to the Majorana wire than it was shown in Fig.~\ref{fig:fig2}.
One can notice that the low-energy spectral function is
restored to the half of its maximum, $\pi\Gamma_\down A_{\downarrow} (0) = 1/2$,
however, this change is not as immediate as it was in the case shown in Fig.~\ref{fig:fig2}(e).
With increasing the spin polarization $p$,
the value of the spectral function at the Fermi level for negligible $V_M$
starts increasing until $p=0.3$, to drop when $p=0.5$.
Such a nonmonotonic behavior can be explained
by realizing that the presence of ferromagnetic leads also affects
the singlet and triplet states of the double quantum dot.
Once the energy of one of the components of the triplet becomes
comparable with the energy of the singlet state, which can happen
for certain value of $p$, the low-energy spectral function 
becomes enhanced due to Kondo processes \cite{Wojcik2015}.
One can also see that the presence of the coupling to topological superconductor
restores the fractional value of 
$\pi\Gamma_\down A_{\downarrow} (0)$ irrespective of $p$.
On the other hand, the spin-up component for large enough
$V_M$ displays the Kondo resonance at the Fermi energy,
which hardly depends on $p$. This is associated with the fact
that, for $\e_2 = -U/3$, coupling to the Majorana wire induces a considerable spin splitting 
of the second quantum dot level, such that the second-stage Kondo screening
becomes then negligible. Note also that the quantum interference
peaks and dips visible in Fig.~\ref{fig:fig2} are now suppressed, 
since the presence of Majorana quasiparticles is mainly revealed
through the splitting $\eM$.
 
Finally, let us discuss the behavior of the spectral functions
when the orbital levels of both quantum dots are set to $\varepsilon_{1} = \varepsilon_{2} = - U/3$.
This case is presented in Fig.~\ref{fig:fig4}.
In this figure one can generally recognize the combination of
the effects that are visible in Figs.~\ref{fig:fig2} and \ref{fig:fig3}.
This is because now we have both the spin splitting
due to the ferromagnetic exchange field as well as the splitting
caused by the coupling to the Majorana wire.
Interestingly, one can see that the behavior of the spin-up
spectral function somewhat resembles the dependence 
presented in the first row of Fig.~\ref{fig:fig2}.
On the other hand, the behavior
of the spin-down spectral function bears some similarity
to the corresponding spectral function shown 
in Fig.~\ref{fig:fig3}. This can be understood by realizing
that the spin-down component is greatly affected
by the presence of Majorana quasiparticles, so 
the results are similar when Majorana splitting is finite,
i.e. in the case of $\e_2 = -U/3$. For the spin-up component
it is the exchange field that is most relevant,
so now the results for $\e_1=-U/3$ bear the corresponding resemblance.
Of course, this analysis is somewhat superficial and
a closer inspection reveals crucial differences.
In particular, it can be seen that for finite $p$
the normalized spectral function for spin-up electrons,
$\pi \Gamma_\uparrow A_\uparrow(\omega)$,
becomes less vulnerable to the influence of the coupling to the topological superconductor,
and is more fixed by the leads spin polarization.
This can be explained by the fact that increasing $p$
induces larger $\exch$, which can partly suppress the first-stage Kondo effect.
Indeed, the value of the spectral function at the Fermi level decreases,
while a pronounced split Kondo peak becomes visible for negative energies.
Moreover, since the quantum interference with the 
Majorana wire is suppressed by finite splitting
of the second dot level due to $\eM$,
changing $V_M$ has a rather weak
influence on $A_\uparrow(\omega)$.
Consequently, now the second-stage Kondo effect cannot be restored,
see Figs.~\ref{fig:fig4}(b-d).

As far as the behavior of the spin-down spectral function is concerned,
one can see much larger dependence on the coupling to the Majorana wire.
For $V_M=0$, increasing $p$ results in a stronger suppression
of $A_\downarrow(0)$, while a split Kondo peak develops
for positive energies. However, even small values of $V_M$
change the picture drastically by bringing the spectral function
at the Fermi energy to the fractional value of 
$\pi \Gamma_\downarrow A_\downarrow(0)=1/2$.
Moreover, while for $p=0.1$ there are clear
maxima (minima) for $\omega<0$ ($\omega>0$)
indicating interference in the spin-down channel,
with increasing $p$ these features become
suppressed. In fact, for $p=0.5$, there is 
a pronounced zero-energy peak of height 
$\pi \Gamma_\downarrow A_\downarrow(0)=1/2$,
the width of which increases with $V_M$, see Fig.~\ref{fig:fig4}(h).

\subsection*{Transport characteristics}

\begin{figure}[t]
	\centering
		\includegraphics[width=\linewidth]{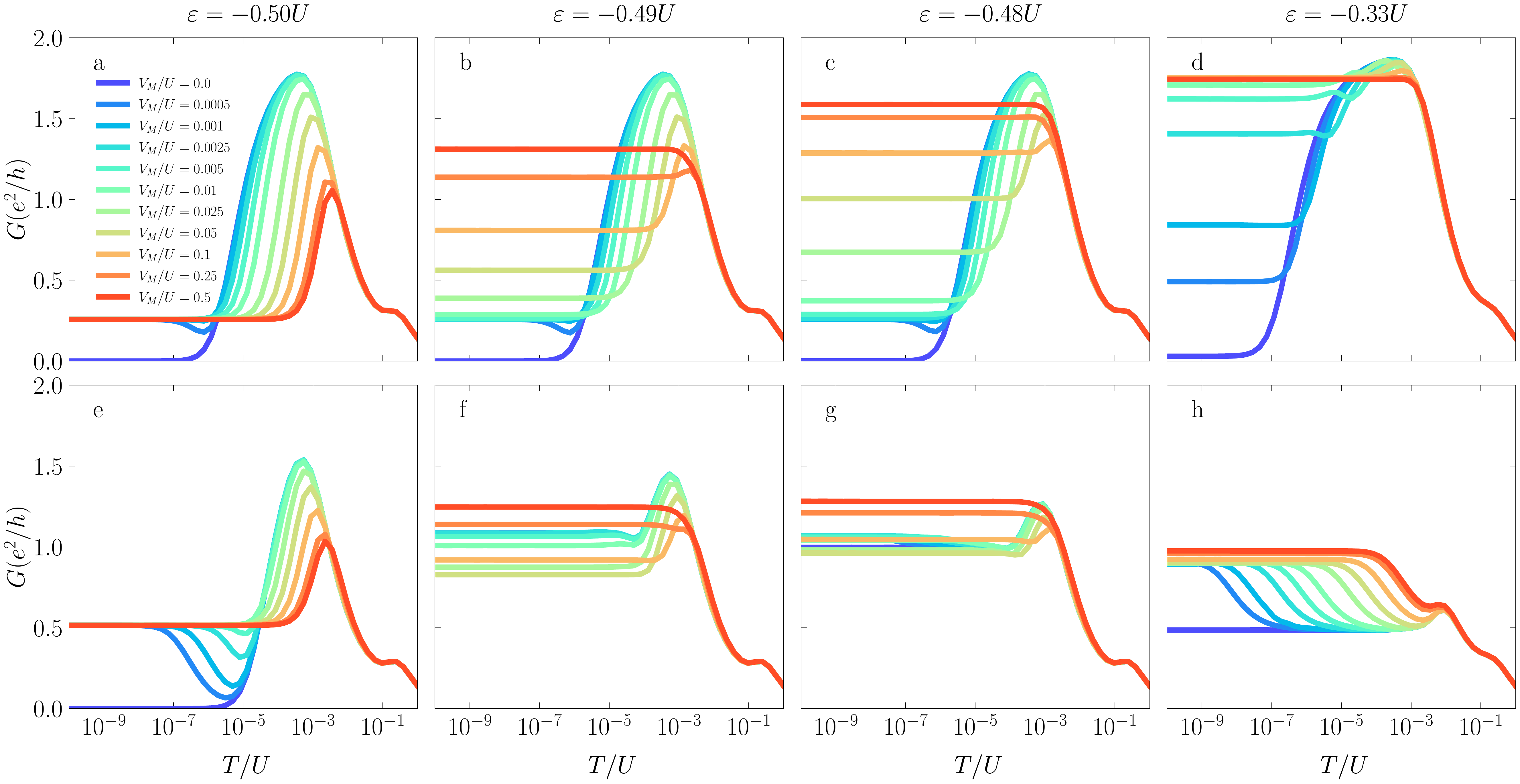}
	\caption{The temperature dependence of 
		the linear conductance for different values of the coupling to Majorana wire, as indicated.
		The first (second) row presents the data for $p=0$ ($p=0.5$),
		while each column corresponds to different value of 
		double dot level positions, $\e\equiv\e_1=\e_2$,
		increasing the detuning from the particle-hole symmetry point,
		while moving to the right.
		The other parameters are the same as in Fig.~\ref{fig:fig4}.}
	\label{fig:fig5}
\end{figure}

Let us now focus on the behavior of the linear conductance of the system.
The linear conductance in the spin channel $\sigma$ can be found from 
\begin{equation}
	G_\sigma = \frac{e^2}{h} \pi\Gamma_\sigma \int d\omega \; A_\sigma (\omega) \left(-\frac{\partial f(\omega)}{\partial \omega}\right),
\end{equation}
where $f(\omega)$ is the Fermi-Dirac distribution function. The total conductance is given by
$G = G_\up + G_\down$, and we can also define the conductance spin polarization as
\begin{equation}
	\Pol = \frac{G_\up - G_\down}{G_\up + G_\down}.
\end{equation}

\subsubsection*{Temperature dependence}

Figure~\ref{fig:fig5} presents the temperature dependence
of the linear conductance for different values of the coupling
to the Majorana wire and for different detuning from the particle-hole
symmetry point, $\e\equiv\e_1 = \e_2 =U/2$.
The first row, for reference, presents the data in the case of $p=0$,
while the second  row displays the results for $p=0.5$.
In the nonmagnetic case, for $\e=-U/2$, one can see that increasing $V_M$
destroys the second stage Kondo screening and leads
to $G(T<T^*)=e^2/2h$. Moreover, one can see that the larger
$V_M$ is, the larger $T^*$ becomes. For finite detuning,
increasing $V_M$ again suppresses the second-stage of the Kondo effect,
however, now the low-temperature conductance acquires a larger value,
which increases with detuning the system more out of the particle-hole symmetry point.
The case of ferromagnetic leads and $\e = -U/2$ is qualitatively similar
to the nonmagnetic case. The only difference is associated
with lower Kondo temperature in the case of finite $p$,
thus the maximum value of the conductance for temperatures
between $T^*$ and $T_K$ is smaller.
On the other hand, when there is a detuning from the particle-hole symmetry point,
the exchange field comes into play suppressing the second-stage of the Kondo effect.
One then finds a finite low-temperature conductance with 
small dependence on the value of $V_M$.
This dependence becomes however considerable
for $\e=-U/3$ when a nonmonotonic dependence of $G(T)$
develops, with an enhancement of conductance around
$T\sim \Gamma_M$, see Fig.~\ref{fig:fig5}(h).

\begin{figure}
	\centering
	\includegraphics[width=1\linewidth]{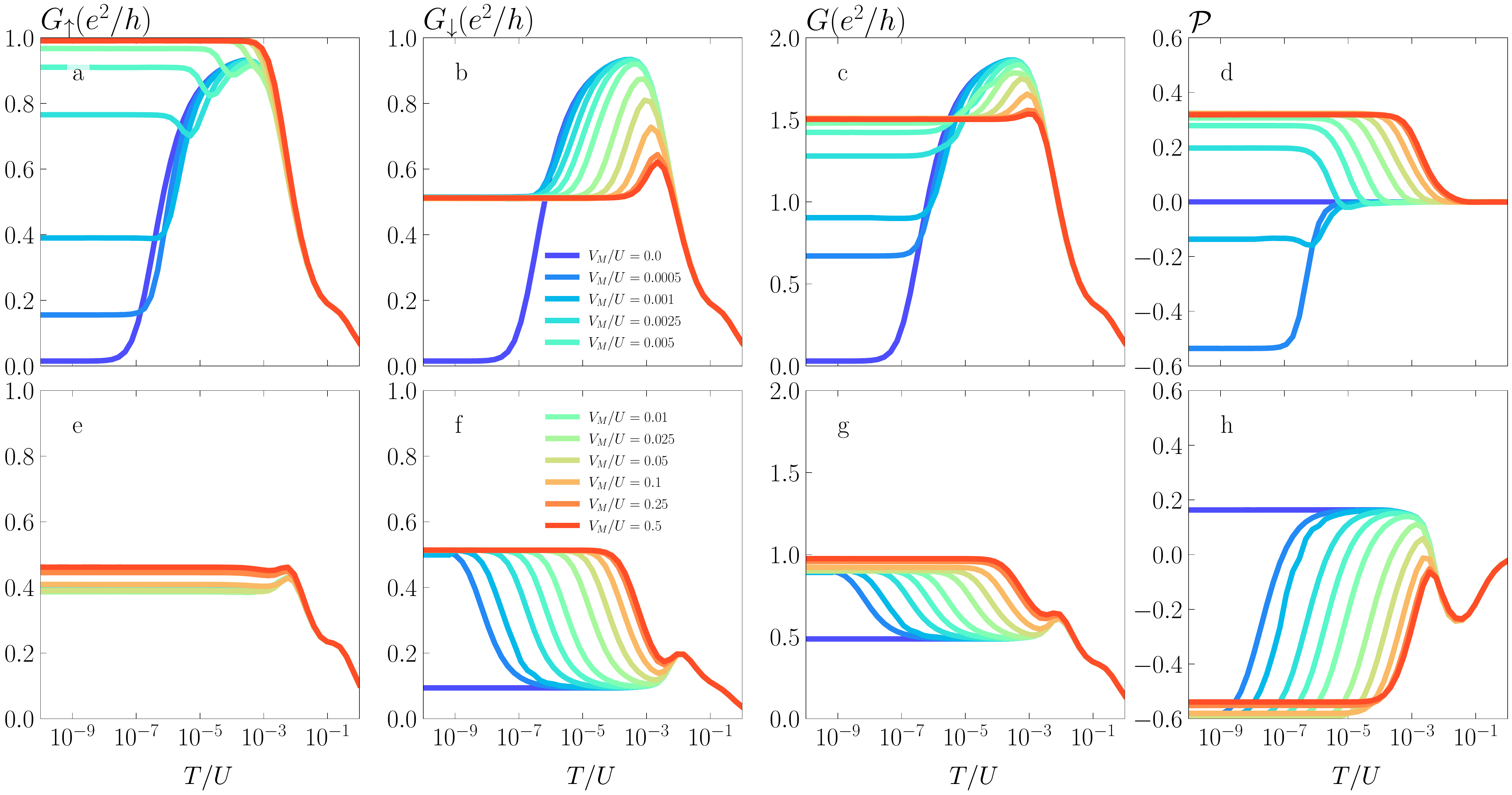}
	\caption{The spin-resolved conductance for the nonmagnetic (first row)
	and ferromagnetic leads with $p=0.5$ (second row) in the case of $\e_1=\e_2=-U/3$.
	The first (second) column shows $G_\up$ ($G_\down$),
	the third column presents $G$, while the last column displays $\Pol$.
	The other parameters are the same as in Fig.~\ref{fig:fig4}.	}
	\label{fig:fig44}
\end{figure}

Let us now inspect the temperature behavior of the spin-resolved
conductance and its polarization for $\e = -U/3$,
again for the nonmagnetic and ferromagnetic lead case.
This is presented in Fig.~\ref{fig:fig44}.
The first (second) column presents $G_{\up}$ ($G_\down$),
the third column displays $G$, while the last column shows $\Pol$ as a function of $T$.
In this figure one can clearly identify various
energy scales determining the system's behavior.
In the case of $p=0$, increasing the coupling to Majorana wire
results in suppression of the second stage of screening,
such that at low temperatures $G_\up$ approaches $e^2/h$, while
$G_\down = e^2/2h$, yielding $G = 3e^2/2h$.
A particular dependence on $V_M$ of the spin components
of the conductance is also revealed in the behavior of $\Pol$.
For low $V_M$, the conductance spin polarization is negative 
at low temperatures, while with increasing $V_M$, it changes sign and becomes positive. 
This is due to the fact that while $G_\down$
increases up to the fractional value of $e^2/2h$ with $V_M$,
$G_\up$ grows up to the maximum value, see the first column of Fig.~\ref{fig:fig44}.
This picture is changed in the case of ferromagnetic leads.
First of all, one can see that $G_\up$ only weakly depends
on the coupling to the Majorana wire. This can be understood
by realizing that it is the spin-down level that is directly
coupled to the Majorana wire and the behavior
of $G_\up$ is mainly determined by the exchange field splitting.
However, this is contrary to the spin-down conductance component,
which exhibits a strong dependence on $V_M$.
One can see that due to the coupling to Majorana wire,
the spin-down conductance becomes enhanced around $T\sim \Gamma_M$.
This behavior is reflected in the conductance spin polarization
which changes sign and becomes negative at low temperatures
in the case of finite coupling to Majorana wire.

\subsubsection*{Interplay of exchange field and Majorana coupling}

\begin{figure}
	\centering
	\includegraphics[width=0.8\linewidth]{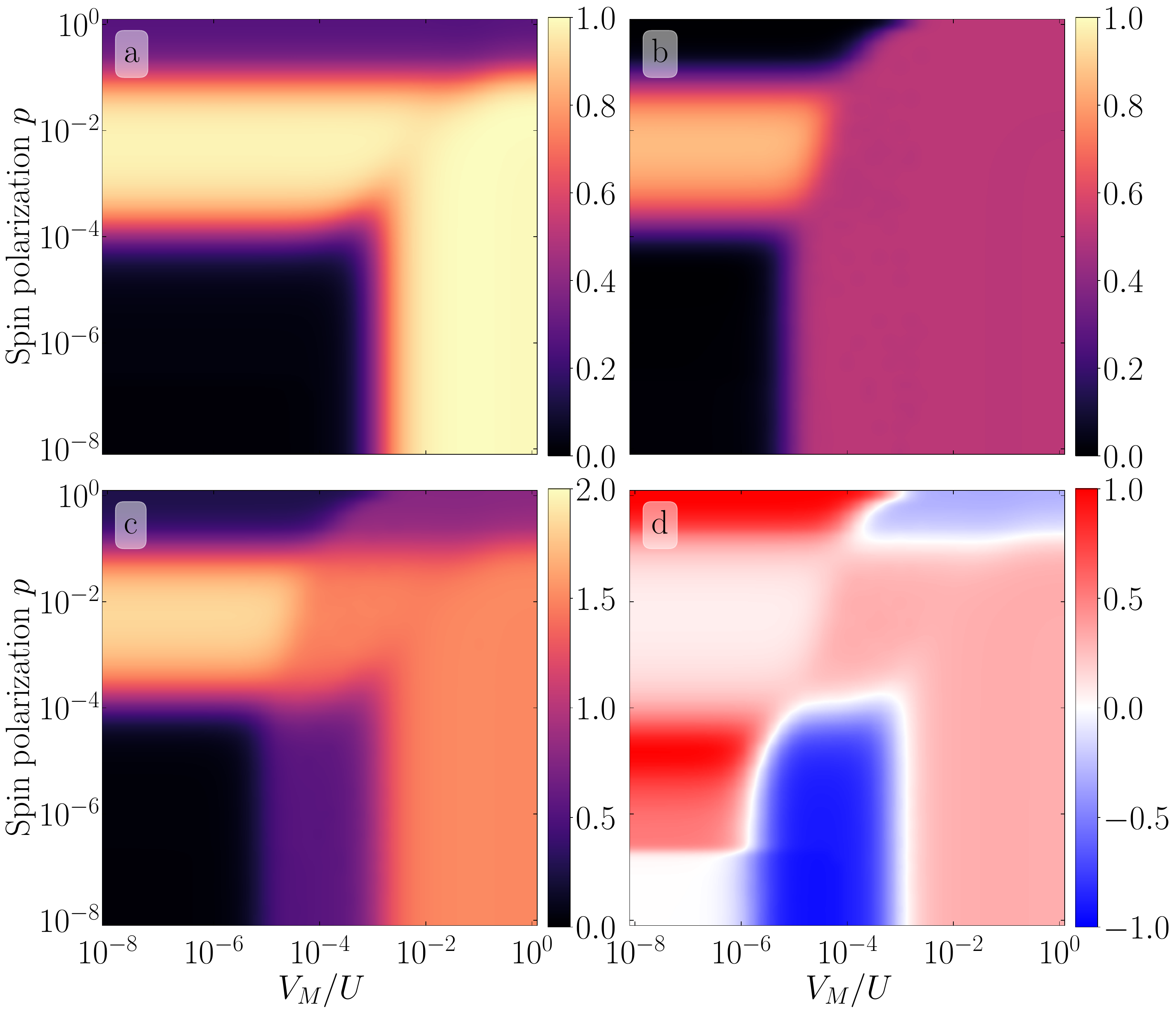}
	\caption{The spin-up (a), spin-down (b) and total (c) 
		linear conductance, as well as the conductance spin polarization (d)
		calculated as a function of the lead spin polarization
		$p$ and the coupling to Majorana wire $V_M$
		assuming low temperature and $\e_1 = \e_2 = -U/3$.
		The other parameters are the same as in Fig.~\ref{fig:fig4}.}
	\label{fig:G2D}
\end{figure}

Figure~\ref{fig:G2D} presents the low-temperature
spin-resolved conductance, $G_\sigma$ and $G$, together with its spin polarization $\Pol$,
plotted as a function of $p$ and $V_M$,
calculated for $\e_1 = \e_2 = -U/3$ and $T = 5 \times 10^{-10}\; U $, which implies $T\ll T^*$.
In this figure one can distinguish significant regimes where the Majorana,
Kondo and exchange field physics interplay.
For negligible spin polarization of the leads $p\to 0$,
one can see that the total conductance exhibits a monotonic increase with $V_M$.
It starts with suppressed conductance $G\approx 0$ for $V_M/U \lesssim 5\times 10^{-6}$
due to the two-stage Kondo effect.
For $5\times 10^{-6} \lesssim V_M /U \lesssim 10^{-3}$,
one finds $G = e^2/2h$, due to the suppressed
second stage of screening in the spin-down channel
yielding $G_\down = e^2/2h$.
Moreover, when $V_M/ U \gtrsim 10^{-3}$, 
large coupling to Majorana wire induces $G_\up = e^2/h$,
which gives rise to $G = 3e^2/2h$.
This behavior is reflected in the dependence of the 
conductance spin polarization, which is generally positive
except for $5\times 10^{-6} \lesssim V_M/U \lesssim 10^{-3}$,
where negative spin polarization develops, $\Pol \approx -1$.

On the other hand, for negligible coupling to Majorana mode, 
increasing lead spin polarization $p$ gives rise to a 
nonmonotonic dependence of the conductance, which is visible in both spin channels.
The mechanism responsible for such nonmonotonicity 
was already explained earlier and is associated with 
spin-splitting of the triplet states caused by the exchange field. 
Moreover, one can see a region of enhanced conductance spin polarization
around $p\approx 2\times 10^{-5}$, as well as for large $p$.
The enhancement for large $p$ is rather intuitive,
since then the spin-up electrons dominate transport,
which results from high spin asymmetry in the couplings.
Such an asymmetry is however not present 
for $p$ as tiny as $p\approx 2\times 10^{-5}$,
and the enhanced spin polarization $\Pol$ is a result
of subtle interplay between the spin splitting caused by the exchange field
and correlations driving the second stage of the Kondo effect.

Interestingly, the above described behavior extends
to finite regions in the $p$ and $V_M$ parameter space presented in Fig.~\ref{fig:G2D}.
For the weak coupling to the Majorana mode and for weakly spin-polarized leads,
the conductance is suppressed, which is the manifestation of the second stage Kondo screening.
This suppression extends up to $p \approx 10^{-4}$
and $V_M/U \approx  5 \times 10^{-6}$.
In this regime, the coupling to Majorana mode
as well as the ferromagnet-induced exchange field 
hardly affect the physics. 
However, when moving towards stronger values of coupling to Majorana mode
and spin polarization, both energy scales start to be relevant.
In this regime one can see the competitive character of both $V_M$ and $p$
affecting the system's transport behavior.
This competition, although hardly visible in the conductance,
is greatly revealed in the dependence of $\Pol$,
which exhibits a sign change along the line $V_M/U \sim p^2$
extending up to $p\approx 10^{-4}$.
Further increasing the spin imbalance in the leads, 
results in $\exch$ winning over spin polarization caused
by the presence of Majorana mode, such that one generally finds $\Pol>0$.
On the other hand, for low $p$, further increasing 
$V_M$ restores the fractional value of the conductance in the spin-down
channel, resulting in maximum negative spin polarization of the conductance,
which however again changes sign and becomes positive once $V_M/U \gtrsim 10^{-3}$.

Note also that Fig.~\ref{fig:G2D}(c) nicely presents 
different fractional values of the conductance extending
from $G\approx 0$, through $G\approx e^2/2h$ and $G\approx 3e^2/2h$,
up to $G\approx 2 e^2/h$. It can be seen that the corresponding conductance plateau regions
not always coincide with the behavior of the conductance spin polarization,
which is shown in Fig.~\ref{fig:G2D}(d). One can then find
regimes of full spin polarization, with spin-up or spin-down
components dominating transport. When the Majorana coupling suppresses the second stage
of the screening in the spin-down channel one finds $\Pol \approx -1$.
However, in the transport regime where the exchange field dominates,
$\Pol$ is enhanced and may reach $\Pol \approx 1$.
On the other hand, for large $V_M$, when the
spin-down conductance takes fractional value and the spin-up
conductance is maximum, one generally finds $\Pol \approx 1/3$, irrespective of $p$.
Finally, it is important to note that the most interesting physics
revealing the interplay between ferromagnetism and coupling to Majorana mode
takes place for low values of $p$ and $V_M$. Such low
values of spin polarization can also result from magnetic impurities or
stray fields present in otherwise nonmagnetic electrodes.

\section*{Conclusion}

We have determined the spin-resolved transport characteristics
of a double quantum dot with ferromagnetic leads side-attached
to topological superconducting nanowire hosting Majorana zero-energy modes.
The model considered is also relevant for systems with 
only one ferromagnetic contact, which could be e.g. a tip of a spin-polarized STM.
We have shown that the behavior of such system is determined by an intricate interplay
between ferromagnet-induced exchange field, quantum interference with
the Majorana wire and the Kondo correlations.
In fact, for the considered setup, the strong electron correlations
can give rise to the first and second-stage Kondo phenomena.
To address this interplay in the most reliable manner,
our investigations have been carried out with the aid of the
non-perturbative numerical renormalization group method.
We have in particular determined the spin-resolved spectral functions of the system,
together with the linear conductance and its spin polarization
for different model parameters, focusing on the transport regimes
where various spin correlation effects come into play.

We have shown that the competing character of splittings
resulting from ferromagnetic contacts and coupling to the Majorana wire
gives rise to unique features in the transport behavior.
First of all, both splittings can suppress the second stage of the Kondo screening
restoring the conductance to a finite value. 
In the case of large coupling to topological superconductor,
this implies a fractional value of the conductance equal to $3e^2/2h$.
On the other hand, in the presence of spin-polarized 
tunneling, the spectral functions exhibit an antisymmetric
resonance at low energies, the position of which depends
on the strength of coupling to the Majorana mode.
The interplay between the corresponding energy scales is also
visible in the behavior of the linear conductance.
While the conductance reaches fractional values due to
the leakage of the Majorana quasiparticle, spin-polarized 
tunneling reduces the conductance and it can destroy
its well-defined fractional character.
Further interesting behavior can be observed
in the temperature dependence of the conductance
spin polarization, which exhibits a nonmonotonic behavior
when tuning the coupling $V_M$ to the Majorana wire, and can
change sign depending on the transport regime.
This sign change is also nicely visible in the 
dependence of conductance spin polarization on both 
the spin imbalance of the leads quantified by $p$
and the coupling to topological superconducting wire.
We show that even very low values of 
electrode spin polarization $p$ can greatly modify the 
system's transport properties giving rise
to almost perfect conductance spin polarization,
whose sign can be tuned depending on both $p$ and $V_M$.
It is important to note that such low imbalance in spin-dependent processes
could be induced e.g. by stray fields, external magnetic field or presence of magnetic impurities. 
Our findings provide thus further unique features
resulting from the presence of Majorana modes
observable in spin-selective transport characteristics.

\section*{Data availability}
The datasets generated and analyzed during the current study are available
from the corresponding author on reasonable request.



\section*{Acknowledgements}
This work was supported by the project No. 2021/41/N/ST3/01885 under the Polish National Science Centre.
We acknowledge the computing time at the Pozna\'n Supercomputing and Networking Center.

\section*{Author contributions statement}
P.M. performed the calculations, prepared the figures and the first draft of the manuscript.
All authors participated in discussions, interpretation of data and finalization of the manuscript.
I.W. coordinated the project.

\section*{Additional information}
Correspondence and requests for materials should be addressed to P.M.

\end{document}